\documentstyle[12pt]{article}
\begin{document}
\makeatletter
\@addtoreset{equation}{section}
\makeatother
\renewcommand{\theequation}{\thesection.\arabic{equation}}
\baselineskip 15pt

\title{\bf A simple note on the nonlocal structure of the PR-box}

\author{GianCarlo Ghirardi\footnote{e-mail: ghirardi@ictp.it}\\
{\small Department of Physics of the University of Trieste, and}\\
{\small the Abdus Salam International Centre for Theoretical Physics,
Trieste, Italy}\\
and\\
\\ Raffaele Romano\footnote{e-mail:rromano@ts.infn.it}\\
{\small Department of Physics of the University of Trieste, and}\\
{\small Fondazione Parisi, Rome, Italy.}}


\maketitle

\begin{abstract}
\noindent In view of the remarkable interest raised by the introduction of the so called PR-box we analyze its nonlocal features from the point of view of Parameter and Outcome Dependence.
\end{abstract}



\section{Introduction} 
 As well known, Bell's inequality implies  that the correlations between the outcomes of measurements performed on  far away quantum systems in entangled states exhibit an irreducible and unavoidable nonlocal nature. Obviously,  such nonlocal correlations might, in principle,  give rise to a conflict with relativistic causality. 
 
 Popescu and Rohlrlich \cite{Popescu} have faced this crucial problem with admirable lucidity and have analyzed in detail the constraints that must be satisfied in order that no conflict with relativistic causality emerges. They have reached quite interesting general conclusions and have also introduced the so-called ``PR-box", a device with two inputs and two outputs which gives rise to  correlations implying what has been called {\it superquantum nonlocality}. The reason for this qualification derives from the fact that the modulus of an appropriate combination of such correlation functions violates not only the classical limit of 2 for it but even the upper limit $2\sqrt{2}$ characteristic of  quantum mechanics.

To illustrate the arguments of these authors we follow their line of thought and we use their notation. Let $A,A',B$ and $B'$ be physical variables taking values +1 and -1, with $A$ and $A'$ referring to measurements on one part of the system by a local observer and $B$ and $B'$ referring to the other part. If we denote as $P_{AB}(a,b)$ the joint probability of obtaining $A=a$ and $B=b$ when both $A$ and $B$ are measured, the correlation $E(A,B)$ of the outcomes is defined as:
 
 \begin{equation}
E(A,B)=P_{AB}(+1,+1)+P_{AB}(-1,-1)  -P_{AB}(+1,-1)-P_{AB}(-1,+1).
\end{equation}

As well known Clauser, Horne, Shimony and Holt \cite{CHSH} have shown, completely in general, that an appropriate combination of such correlations (with the variables $A,A',B,B'$   arbitrarily chosen)   satisfies, for all local theories, the inequality:

\begin{equation}
\vert E(A,B)+E(A,B')+E(A',B)-E(A',B')\vert\leq{2}.
\end{equation}

On the other hand, in the quantum case, when consideration is given to two far away spin $1/2$ particles in the singlet state and measurements of the spin components along appropriate directions are performed, the correlations $E_{Q}(A,B)\equiv \langle\Psi\vert A^{(1)}\otimes B^{(2)}\vert\Psi\rangle$ violate, for appropriate choices of $A,A',B,B'$, the above inequality. Actually, Bell \cite{Bell} has derived his celebrated inequality:

\begin{equation}
\vert E_{Q}(A,B)+E_{Q}(A,B')+E_{Q}(A',B)-E_{Q}(A',B')\vert\leq{2\sqrt{2}},
\end{equation}

\noindent and, as well known, the right hand side upper bound can  actually be reached for appropriate choices of the observables appearing in it.

The authors of ref.\cite{Popescu} have investigated whether the request that the hypothetical general nonlocal theory one is envisaging respects relativistic causality might be responsible for the precise value of Bell's upper bound. The question is interesting since, at first sight, one might expect that the above combination of correlations reaches the value 4, which is attained when the first three terms take the value +1 and the last the value -1. The extremely interesting result of ref.\cite{Popescu} is the proof that, in principle, a nonlocal theory respecting relativistic causality and yielding  a value greater than the upper bound is possible (we will call any theory exhibiting such a feature a {\it superquantum nonlocal theory}). Secondly, by  resorting to the smart consideration of the PR-box, the authors have identified a specific family of correlations which, without conflicting with relativistic causality, actually reach the theoretical upper bound of 4. 

At this stage it is interesting to mention that other conceptual analysis  \cite{Jarrett, Suppes, Shimony} concerning the locality issue have led to the conclusion that nonlocality in Bell's sense amounts to the logical conjunction of two other requests which have been named  {\it Locality} and {\it Completeness}, respectively,  by Jarret  \cite {Jarrett} and {\it Parameter Independence (PI)} and {\it Outcome Independence (OI)}, by Shimony  \cite{Shimony}. The distinction involved is quite elementary. Let us call $P(A=a|x,y),\;P(B=b|x,y),\;P(A=a,B=b|x,y), $ and $P(A=a|x,y;B=b)$ etc., the single and joint, conditional and unconditional probabilities of the outcomes $(a,b)$ for the inputs (settings) $(x,y)$, and let us recall the relation for conditional probabilities:

\begin{equation}
P(A=a,B=b|x,y)=P(A=a|x,y;B=b)\cdot P(B=b|x,y).
\end{equation}

\noindent If one assumes {\it Completeness $\equiv$ Outcome Independence}:

\begin{eqnarray}
P(A=a|x,y;B=b)&=&P(A=a|x,y),\\ \nonumber P(B=b|x,y;A=a)&=&P(B=b|x,y),
\end{eqnarray}

\noindent and {\it Locality $\equiv$ Parameter Independence}:

\begin{equation}
P(A=a|x,y)=P(A=a,|x),\;\;P(B=b|x,y)=P(B=b|y),
\end{equation}

\noindent one gets
\begin{equation}
P(A=a,B=b|x,y)=P(A=a,|x)\cdot P(B=b|y),
\end{equation}

\noindent i.e. Bell's locality request. On the other hand it is trivial to go the other way around showing that this last condition implies both {\it Locality} and {\it Completeness}.

\vspace {0.3cm}

\section {Analysis of the nonlocal features of the PR-box} 
The characterization of the PR-box is quite simple. It is represented by the following relation between the inputs $(x,y)$ and the outcomes $(a,b)$, each of which is assumed to take only the values $\{0,1\}$:

\begin{equation}
a+b=xy\;\; mod \;2.
\end{equation}

We will analize the PR-box under two possible formulation of its working, the first one (Case 1) being given simply by Eq.(2.1), the second one (Case 2) being enriched by the introduction of a deterministic hidden variable description of the inputs-outputs relations.
\vspace {0.3cm}

\subsection {Case 1} When one takes into account only the relation (2.1) one can argue in the following way.
\begin{itemize}
\item Let us consider the system $A$ and suppose that its input $x$ is known. Then, if $x=0 \rightarrow xy=0 \rightarrow (a=0, b=0)\vee (a=1,b=1)$. So, the outcome $a$ that Alice (at $A$) will get once she knows her setting $x=0$, depends on the outcome, 0 or 1, that Bob (at $B$) has obtained. Outcome Independence is violated. On the contrary, if $x=1$ two cases are possible: either $y=0$ and we are back to the previous situation, or $y=1 \rightarrow (a=1, b=0)\vee (a=0,b=1)$. In this case, both knowledge of $b$ and of $y$ are necessary to know $a$ uniquely. So, given the input at $A$,  the corresponding output depends, in general,  both from $y$ and $b$: the theory violates both PI as well as OI.
\item Completely analogous considerations hold for the setting and the outcome at B.
\item Another way of looking at the problem derives from looking at the product $ab$ of the outcomes. In the case in which both settings at $A$ and at $B$ are given, there are two possibilities: if $xy=0$ we know that for sure $a=b$ but we do not know whether they take the value $0$ or the value $1$. Alternatively, if $xy=1$ we know that one between a and b takes the value $0$ and one the value $1$, but, once more, we do not know the actual value of them. Specification of both settings does not determine the outcomes. Some further knowledge is necessary.
\end{itemize}

\subsection {Case 2} Suppose now we consider a hidden variable model characterized by a  variable $\lambda$ which also can take the values $\{0,1\}$ and, to be completely general, let us assume that the probabilities of its two values are given by $P^{(\lambda)}(0)$ and $P^{(\lambda)}(1)$, with, obviously, $P^{(\lambda)}(0)+P^{(\lambda)}(1)=1$. 

The model is defined by the assumption that, for any given setting $x$ for $A$ and/or $y$ for $B$, the assignment  of $\lambda$ determines the  outcome(s) according to the following rules: 

\begin{eqnarray}
a &=& (x+\lambda)\;\;  mod \;2 \nonumber \\
b &=& (x+\lambda-xy)\;\; mod \;2. 
\end{eqnarray}

\noindent Note that the model is manifestly nonlocal since the value of the outcome $b$ besides depending on the value of the hidden variable $\lambda$ and of the associated input $y$ depends also on the input $x$.

In accordance with the above rules  the assignments of the outputs  (once the settings and the hidden variable are given) is the one exhibited in the following  table:

\begin{table}[htdp]
\caption{Hidden Variable model outcomes as functions of the inputs and of $\lambda$.}
\begin{center}
\begin{tabular}{c|c|c||cc}
x & y & $\lambda$ & a & b \\ [0.5ex]  \hline
0 & 0 & 0 & 0 & 0 \\
0 & 0 & 1 & 1 & 1 \\
1 & 0 & 0 & 1 & 1 \\
1 & 0 & 1 & 0 & 0 \\
0 & 1 & 0 & 0 & 0 \\
0 & 1 & 1 & 1 & 1 \\
1 & 1 & 0 & 1 & 0 \\
1 & 1 & 1 & 0 & 1 

\end{tabular}
\end{center}
\label{tav}
\end{table}%

\noindent Looking at the table one immediately checks that the basic relation characterizing the inputs and outputs of the PR-Box: $a+b=xy\;\;mod \;2$ is satisfied. 

It is important to remark that the outputs concerning  the two parties depend on the inputs of the other party (nonlocality) but  not  on the associated output. For what concerns us here we can then claim that  the model exhibits Parameter (as any deterministic nonlocal hidden variable model) but not Outcome dependence.

\vspace {0.3cm}

\section {The superquantum character of the PR-box } 
We have repeatedly claimed that the PR-Box gives rise to superquantum nonlocal correlations. To show this, by following the procedure of  ref.\cite{Cerf}, it is useful to introduce new outcomes $a'$ and $b'$, which are simply related to $a$ and $b$ in such a way that their possible values are $\{-1,+1\}$, as in Eqs. (1.1)-(1.3). We then put:

\begin{equation}
a'=1-2a;\;\;\;b'=1-2b,
\end{equation}
and we consider the quantities $E_{PR-HV}(x,y)=P_{xy}(+1,+1)+P_{xy}(-1,-1)-P_{xy}(+1,-1)-P_{xy}(-1,+1)$, which are easily calculated. Just to give an example, since, as immediately seen from  the table when $x=0$ and $y=0$  it turns out that for any value of $\lambda$, $a=b$ (implying $a'=b'$) one gets $E_{PR-HV}(0,0)= P^{(\lambda)}(0)+P^{(\lambda)}(1)=1$. Just in the same way one reaches the same conclusion for $x=0$ and $y=1$ and for $x=1$ and $y=0$. On the contrary, when $x=1$ and $y=1$ the outcomes $a$ and $b$ are different so that only $P_{11}(+1,-1)$ and $P_{11}(-1,+1)$ contribute to $E_{PR-HV}(1,1)$. But the sum of these probabilities  equals the probability  that $\lambda$ takes the value $0$ or $1$, which is 1. Accordingly  also $-E_{PR-HV}(1,1)=P_{11}(+1,-1)+P_{11}(-1,+1)$ takes  the value $+1$ and the general combination expressing the violation of locality reads:

\begin{equation}
E_{PR-HV}(0,0)+E_{PR-HV}(1,0)+E_{PR-HV}(0,1)-E_{PR-HV}(1,1)=4,
\end{equation}

\noindent i.e. the considered combination actually saturates its upper bound.

\vspace {0.3cm}

\section {Concluding remarks} We have analyzed the superquantum nonlocal structure which characterizes the PR-box and we have shown that, if one considers only the general formal structure characterizing the model embodied by Eq.(2.1), the ensuing nonlocal theory exhibits both Parameter Dependence and Outcome Dependence. On the contrary, it is quite simple to account for the working of the box in terms of a deterministic hidden variable theory\footnote{Actually the  hidden variables theories we have considered represent already a continuous family because all of them reproduce the functioning of the PR-Box, independently of the explicit values chosen for the hidden variable distribution, provided their probabilities satisfy the necessary request $P^{(\lambda)}(0)+P^{(\lambda)}(1)=1$} . In such a case, as all nonlocal deterministic models, the PR-box turns out to violate the locality request because of the violation of Parameter Independence.


\end{document}